\pdfoutput=1

\documentclass[12pt,a4paper]{article}

\usepackage{ifthen} 

\usepackage{cleveref}
\crefname{chapter}{Ch.\@}{Chs.\@}
\crefname{section}{Sec.\@}{Secs.\@}
\crefname{subsection}{Sec.\@}{Secs.\@}
\crefname{appendix}{Appendix\@}{Appendices\@}
\crefname{figure}{Fig.\@}{Figs.\@}
\crefname{table}{Table\@}{Tables\@}
\crefname{equation}{Eq.\@}{Eqs.\@}
\usepackage{booktabs}
\usepackage{siunitx}
\usepackage{tikz}
\usetikzlibrary{external,calc,shapes,decorations,decorations.text,mindmap,shadings,patterns,matrix,arrows,intersections,automata,backgrounds}
\usepackage{subfig}
\newboolean{pdflatex}
\setboolean{pdflatex}{true} 

\newboolean{articletitles}
\setboolean{articletitles}{true} 

\newboolean{uprightparticles}
\setboolean{uprightparticles}{false} 

\newboolean{inbibliography}
\setboolean{inbibliography}{false} 

\def\paperauthors{The LHCb collaboration} 
\def\paperasciititle{Measurement of CP violation in B0 -> JPsi K0S and B0 -> Psi(2S) K0S decays} 
\def\papertitle{Measurement of \CP violation in \BdToJpsiKS and \BdToPsiTwoSKS decays} 
\def\paperkeywords{{High Energy Physics}, {LHCb}} 
\def\papercopyright{CERN on behalf of the LHCb collaboration}
\def\paperlicence{CC-BY-4.0}
\def\paperlicenceurl{https://creativecommons.org/licenses/by/4.0/}

\usepackage[top=1in, bottom=1.25in, left=1in, right=1in]{geometry}

\usepackage{microtype}
\usepackage{lineno}  
\usepackage{xspace} 
\usepackage{caption} 

\usepackage{graphicx}  
\usepackage{color}
\usepackage{colortbl}
\graphicspath{{./figs/}} 

\usepackage{siunitx}
\usepackage{amsmath} 
\usepackage{amssymb}
\usepackage{amsfonts}
\usepackage{upgreek} 
\usepackage{fancyvrb} 

\newcommand*\patchAmsMathEnvironmentForLineno[1]{%
\expandafter\let\csname old#1\expandafter\endcsname\csname #1\endcsname
\expandafter\let\csname oldend#1\expandafter\endcsname\csname
end#1\endcsname
 \renewenvironment{#1}%
   {\linenomath\csname old#1\endcsname}%
   {\csname oldend#1\endcsname\endlinenomath}%
}
\newcommand*\patchBothAmsMathEnvironmentsForLineno[1]{%
  \patchAmsMathEnvironmentForLineno{#1}%
  \patchAmsMathEnvironmentForLineno{#1*}%
}
\AtBeginDocument{%
\patchBothAmsMathEnvironmentsForLineno{equation}%
\patchBothAmsMathEnvironmentsForLineno{align}%
\patchBothAmsMathEnvironmentsForLineno{flalign}%
\patchBothAmsMathEnvironmentsForLineno{alignat}%
\patchBothAmsMathEnvironmentsForLineno{gather}%
\patchBothAmsMathEnvironmentsForLineno{multline}%
\patchBothAmsMathEnvironmentsForLineno{eqnarray}%
}

\usepackage{hyperxmp}

\usepackage[pdftex,
            pdfauthor={\paperauthors},
            pdftitle={\paperasciititle},
            pdfkeywords={\paperkeywords},
            pdfcopyright={Copyright (C) \papercopyright},
            pdflicenseurl={\paperlicenceurl}]{hyperref}

\usepackage[all]{hypcap} 

\usepackage{xspace} 
\usepackage{upgreek}

\def\lhcb {\mbox{LHCb}\xspace}

\def\MagUp {\mbox{\em Mag\kern -0.05em Up}\xspace}

\ifthenelse{\boolean{uprightparticles}}%
{

 \def\Peta        {\ensuremath{\upeta}\xspace}

 \def\Ppi         {\ensuremath{\uppi}\xspace}

 \def\Ppsi        {\ensuremath{\uppsi}\xspace}

 \def\PDelta      {\ensuremath{\Delta}\xspace}                 
 \def\PXi      {\ensuremath{\Xi}\xspace}                 
 \def\PLambda      {\ensuremath{\Lambda}\xspace}                 
 \def\PSigma      {\ensuremath{\Sigma}\xspace}                 
 \def\POmega      {\ensuremath{\Omega}\xspace}                 
 \def\PUpsilon      {\ensuremath{\Upsilon}\xspace}                 
 
 \def\PB      {\ensuremath{\mathrm{B}}\xspace}                 
                  
 \def\PD      {\ensuremath{\mathrm{D}}\xspace}

 \def\PJ      {\ensuremath{\mathrm{J}}\xspace}                 
 \def\PK      {\ensuremath{\mathrm{K}}\xspace}

 \def\Pb      {\ensuremath{\mathrm{b}}\xspace}                 
 \def\Pc      {\ensuremath{\mathrm{c}}\xspace}                 
 \def\Pd      {\ensuremath{\mathrm{d}}\xspace}

 \def\Pi      {\ensuremath{\mathrm{i}}\xspace}

 \def\Ps      {\ensuremath{\mathrm{s}}\xspace}                 
 \def\Pt      {\ensuremath{\mathrm{t}}\xspace}

}
{

 \def\Peta        {\ensuremath{\eta}\xspace}

 \def\Ppi         {\ensuremath{\pi}\xspace}

 \def\Ppsi        {\ensuremath{\psi}\xspace}                 
                  
 \mathchardef\PDelta="7101
 \mathchardef\PXi="7104
 \mathchardef\PLambda="7103
 \mathchardef\PSigma="7106
 \mathchardef\POmega="710A
 \mathchardef\PUpsilon="7107
                  
 \def\PB      {\ensuremath{B}\xspace}                 
                  
 \def\PD      {\ensuremath{D}\xspace}

 \def\PJ      {\ensuremath{J}\xspace}                 
 \def\PK      {\ensuremath{K}\xspace}

 \def\Pb      {\ensuremath{b}\xspace}                 
 \def\Pc      {\ensuremath{c}\xspace}                 
 \def\Pd      {\ensuremath{d}\xspace}

 \def\Pi      {\ensuremath{i}\xspace}

 \def\Ps      {\ensuremath{s}\xspace}                 
 \def\Pt      {\ensuremath{t}\xspace}

}

\makeatletter
\ifcase \@ptsize \relax
  \newcommand{\miniscule}{\@setfontsize\miniscule{4}{5}}
\or
  \newcommand{\miniscule}{\@setfontsize\miniscule{5}{6}}
\or
  \newcommand{\miniscule}{\@setfontsize\miniscule{5}{6}}
\fi
\makeatother

\DeclareRobustCommand{\optbar}[1]{\shortstack{{\miniscule (\rule[.5ex]{1.25em}{.18mm})}
  \\ [-.7ex] $#1$}}

\def\dquark    {{\ensuremath{\Pd}}\xspace}

\def\squark    {{\ensuremath{\Ps}}\xspace}

\def\cquark    {{\ensuremath{\Pc}}\xspace}
\def\cquarkbar {{\ensuremath{\overline \cquark}}\xspace}
\def\ccbar     {{\ensuremath{\cquark\cquarkbar}}\xspace}
\def\bquark    {{\ensuremath{\Pb}}\xspace}
\def\bquarkbar {{\ensuremath{\overline \bquark}}\xspace}

\def\tquark    {{\ensuremath{\Pt}}\xspace}

\def\pion   {{\ensuremath{\Ppi}}\xspace}

\def\pip    {{\ensuremath{\pion^+}}\xspace}
\def\pim    {{\ensuremath{\pion^-}}\xspace}

\def\kaon    {{\ensuremath{\PK}}\xspace}
  \def\Kbar    {{\kern 0.2em\overline{\kern -0.2em \PK}{}}\xspace}

\def\KorKbar    {\kern 0.18em\optbar{\kern -0.18em K}{}\xspace}
\def\Kz      {{\ensuremath{\kaon^0}}\xspace}
\def\Kzb     {{\ensuremath{\Kbar{}^0}}\xspace}
\def\Kp      {{\ensuremath{\kaon^+}}\xspace}

\def\KS      {{\ensuremath{\kaon^0_{\mathrm{ \scriptscriptstyle S}}}}\xspace}

\def\Kstarz  {{\ensuremath{\kaon^{*0}}}\xspace}

  \def\Dbar    {{\kern 0.2em\overline{\kern -0.2em \PD}{}}\xspace}

\def\DorDbar    {\kern 0.18em\optbar{\kern -0.18em D}{}\xspace}

\def\B       {{\ensuremath{\PB}}\xspace}
\def\Bbar    {{\ensuremath{\kern 0.18em\overline{\kern -0.18em \PB}{}}}\xspace}

\def\BorBbar    {\kern 0.18em\optbar{\kern -0.18em B}{}\xspace}
\def\Bz      {{\ensuremath{\B^0}}\xspace}
\def\Bzb     {{\ensuremath{\Bbar{}^0}}\xspace}
\def\Bu      {{\ensuremath{\B^+}}\xspace}

\def\Bd      {{\ensuremath{\B^0}}\xspace}
\def\Bs      {{\ensuremath{\B^0_\squark}}\xspace}

\def\jpsi     {{\ensuremath{{\PJ\mskip -3mu/\mskip -2mu\Ppsi\mskip 2mu}}}\xspace}
\def\psitwos  {{\ensuremath{\Ppsi{(2S)}}}\xspace}

\def\etac     {{\ensuremath{\Peta_\cquark}}\xspace}

  \def\Y#1S{\ensuremath{\PUpsilon{(#1S)}}\xspace}

\def\Lz          {{\ensuremath{\PLambda}}\xspace}
\def\Lbar        {{\ensuremath{\kern 0.1em\overline{\kern -0.1em\PLambda}}}\xspace}
\def\LorLbar    {\kern 0.18em\optbar{\kern -0.18em \PLambda}{}\xspace}

\def\Lb      {{\ensuremath{\Lz^0_\bquark}}\xspace}

\newcommand{\decay}[2]{\ensuremath{#1\!\to #2}\xspace}         

\def\to                 {\ensuremath{\rightarrow}\xspace}

\def\CP                {{\ensuremath{C\!P}}\xspace}

\def\Vcbs  {{\ensuremath{V_{\cquark\bquark}^\ast}}\xspace}
\def\Vtbs  {{\ensuremath{V_{\tquark\bquark}^\ast}}\xspace}

\newcommand{\dm}{{\ensuremath{\Delta m}}\xspace}

\newcommand{\DG}{{\ensuremath{\Delta\Gamma}}\xspace}

\def\AT#1     {\ensuremath{A_{\mathrm{T}}^{#1}}\xspace}           

\def\C#1      {\ensuremath{\mathcal{C}_{#1}}\xspace}                       
\def\Cp#1     {\ensuremath{\mathcal{C}_{#1}^{'}}\xspace}                    
\def\Ceff#1   {\ensuremath{\mathcal{C}_{#1}^{\mathrm{(eff)}}}\xspace}        
\def\Cpeff#1  {\ensuremath{\mathcal{C}_{#1}^{'\mathrm{(eff)}}}\xspace}       
\def\Ope#1    {\ensuremath{\mathcal{O}_{#1}}\xspace}                       
\def\Opep#1   {\ensuremath{\mathcal{O}_{#1}^{'}}\xspace}                    

\newcommand{\tev}{\ifthenelse{\boolean{inbibliography}}{\ensuremath{~T\kern -0.05em eV}}{\ensuremath{\mathrm{\,Te\kern -0.1em V}}}\xspace}
\newcommand{\gev}{\ensuremath{\mathrm{\,Ge\kern -0.1em V}}\xspace}
\newcommand{\mev}{\ensuremath{\mathrm{\,Me\kern -0.1em V}}\xspace}
\newcommand{\kev}{\ensuremath{\mathrm{\,ke\kern -0.1em V}}\xspace}
\newcommand{\ev}{\ensuremath{\mathrm{\,e\kern -0.1em V}}\xspace}
\newcommand{\gevc}{\ensuremath{{\mathrm{\,Ge\kern -0.1em V\!/}c}}\xspace}
\newcommand{\mevc}{\ensuremath{{\mathrm{\,Me\kern -0.1em V\!/}c}}\xspace}
\newcommand{\gevcc}{\ensuremath{{\mathrm{\,Ge\kern -0.1em V\!/}c^2}}\xspace}
\newcommand{\gevgevcccc}{\ensuremath{{\mathrm{\,Ge\kern -0.1em V^2\!/}c^4}}\xspace}
\newcommand{\mevcc}{\ensuremath{{\mathrm{\,Me\kern -0.1em V\!/}c^2}}\xspace}

\def\mum  {\ensuremath{{\,\upmu\mathrm{m}}}\xspace}

\def\barn{\ensuremath{\mathrm{ \,b}}\xspace}
\def\nb {\ensuremath{\mathrm{ \,nb}}\xspace}

\def\pb {\ensuremath{\mathrm{ \,pb}}\xspace}

\def\fb   {\ensuremath{\mbox{\,fb}}\xspace}

\def\ab   {\ensuremath{\mbox{\,ab}}\xspace}

\def\ps   {\ensuremath{{\mathrm{ \,ps}}}\xspace}

\def\invps{\ensuremath{{\mathrm{ \,ps^{-1}}}}\xspace}

\newcommand{\chisq}{\ensuremath{\chi^2}\xspace}

\newcommand{\chisqip}{\ensuremath{\chi^2_{\text{IP}}}\xspace}

\newcommand{\chisqvtx}{\ensuremath{\chi^2_{\text{vtx}}}\xspace}

\def\gsim{{~\raise.15em\hbox{$>$}\kern-.85em
          \lower.35em\hbox{$\sim$}~}\xspace}
\def\lsim{{~\raise.15em\hbox{$<$}\kern-.85em
          \lower.35em\hbox{$\sim$}~}\xspace}

\def\sPlot{\mbox{\em sPlot}\xspace}

\def\ptot       {\mbox{$p$}\xspace}
\def\pt         {\mbox{$p_{\mathrm{ T}}$}\xspace}

\def\mrad{\ensuremath{\mathrm{ \,mrad}}\xspace}
\def\rad{\ensuremath{\mathrm{ \,rad}}\xspace}

\def\tell1  {TELL1\xspace}
\def\ukl1   {UKL1\xspace}

\newcommand{\ie}{\mbox{\itshape i.e.}\xspace}

\newcommand{\etc}{\mbox{\itshape etc.}\xspace}

\newcommand{\clight}{\ensuremath{c}\xspace}

\newcommand{\Vcdnew}{{\ensuremath{V^{\phantom{*}}_{\cquark\dquark}}}\xspace}
\newcommand{\Vtdnew}{{\ensuremath{V^{\phantom{*}}_{\tquark\dquark}}}\xspace}

\SaveVerb{StripDetached}=StrippingBetaSBd2JpsiKsDetached=

\newcommand{\BdToJpsiKst}{\decay{\Bd}{\jpsi\Kstarz}}

\newcommand{\psitwoS}{\ensuremath{\psi(2S)}\xspace}

\newcommand{\BdToPsiTwoSKS}{\decay{\Bd}{\psi(2S)\KS}}
\newcommand{\BdToPsiTwoSKSBfac}{\decay{\Bd}{\psi(2S)\KS}}

\SaveVerb{StripmumuPrescaled}=StrippingBetaSPsi2SMuMu_Bd2Psi2SKsMuMuPrescaled=
\SaveVerb{Reco14}=Reco14=

\newcommand{\Jpsi}{\jpsi}
\newcommand{\BdToJpsiKS}{\decay{\Bd}{\jpsi\KS}}

\newcommand{\BuToJpsiK}{\decay{\Bu}{\jpsi\Kp}}

\newcommand{\sintwobeta}{\ensuremath{\sin 2\beta}\xspace}

\SaveVerb{TriggerRequirement}=(Hlt1DiMuonHighMass || Hlt1TrackMuon) && Hlt2DiMuonDetachedJPsi=
\SaveVerb{AlmostUnbiased}=Hlt1DiMuonHighMass && Hlt2DiMuonDetachedJPsi=
\SaveVerb{AlmostUnbiasedEnumerator}=Hlt1DiMuonHighMass && Hlt2DiMuonDetachedJPsi && Hlt2DiMuonJPsi=
\SaveVerb{ExclusivelyBiased}=Hlt1TrackMuon && Hlt2DiMuonDetachedJPsi && !Hlt1DiMuonHighMass=
\SaveVerb{ExclusivelyUnbiased}=Hlt1DiMuonHighMass && Hlt2DiMuonJPsi=
\SaveVerb{2011TriggerRequirement}=Hlt1DiMuonHighMass && Hlt2DiMuonDetachedJPsi=
\SaveVerb{TriggerAllOrRequirement}=Hlt1DiMuonHighMass || Hlt1TrackMuon || Hlt2DiMuonDetachedJPsi || Hlt2DiMuonJPsi=
\SaveVerb{L0}=L0=
\SaveVerb{Hlt1}=Hlt1=
\SaveVerb{Hlt2}=Hlt2=
\SaveVerb{Hlt1u}=J_psi_1S_Hlt1DiMuonHighMassDecision_TOS=
\SaveVerb{Hlt1b}=J_psi_1S_Hlt1TrackMuonDecision_TOS=
\SaveVerb{Hlt2u}=J_psi_1S_Hlt2DiMuonJPsiDecision_TOS=
\SaveVerb{Hlt2b}=J_psi_1S_Hlt2DiMuonDetachedJPsiDecision_TOS=
\SaveVerb{ShortHlt1u}=Hlt1DiMuonHighMass=
\SaveVerb{ShortHlt1b}=Hlt1TrackMuon=
\SaveVerb{ShortHlt2u}=Hlt2DiMuonJPsi=
\SaveVerb{ShortHlt2b}=Hlt2DiMuonDetachedJPsi=
\SaveVerb{And}= && =
\SaveVerb{Or}= || =
\SaveVerb{StripDetached}=StrippingBetaSBd2JpsiKsDetached=
\SaveVerb{StripPrescaled}=StrippingBetaSBd2JpsiKsPrescaled=

\newcommand{\dOS}{\ensuremath{d'_{\text{OS}}}\xspace}
\newcommand{\dSS}{\ensuremath{d'_{\text{SS}}}\xspace}
\newcommand{\etaOS}{\ensuremath{\eta_{\text{OS}}}\xspace}
\newcommand{\etaSS}{\ensuremath{\eta_{\text{SS}}}\xspace}

\newcommand{\prodasym}[1]{\ensuremath{A_\text{P}^{#1}}\xspace}

\DeclareSIUnit\clight{\ensuremath{\mathit{c}}}

\DeclareSIUnit\rad{\radian}
\DeclareSIUnit\mrad{\milli\rad} 

\DeclareSIUnit\micron{\micro\metre} 

\DeclareSIUnit\nanobarn{\nano\barn} 
\DeclareSIUnit\picobarn{\pico\barn} 
\DeclareSIUnit\femtobarn{\femto\barn} 
\DeclareSIUnit\attobarn{\atto\barn}
\DeclareSIUnit\nb{\nano\barn} 
\DeclareSIUnit\pb{\pico\barn} 
\DeclareSIUnit\fb{\femto\barn} 
\DeclareSIUnit\ab{\atto\barn} 
\DeclareSIUnit\zb{\zepto\barn} 
\DeclareSIUnit\yb{\yocto\barn}

\DeclareSIUnit[per-mode=symbol]\eVc{\eV\per\clight}
\DeclareSIUnit[per-mode=symbol]\keVc{\kilo\eV\per\clight}
\DeclareSIUnit[per-mode=symbol]\MeVc{\mega\eV\per\clight}
\DeclareSIUnit[per-mode=symbol]\GeVc{\giga\eV\per\clight}
\DeclareSIUnit[per-mode=symbol]\TeVc{\tera\eV\per\clight}

\DeclareSIUnit[per-mode=symbol]\eVcc{\eV\per\square\clight}
\DeclareSIUnit[per-mode=symbol]\keVcc{\kilo\eV\per\square\clight}
\DeclareSIUnit[per-mode=symbol]\MeVcc{\mega\eV\per\square\clight}
\DeclareSIUnit[per-mode=symbol]\GeVcc{\giga\eV\per\square\clight}
\DeclareSIUnit[per-mode=symbol]\TeVcc{\tera\eV\per\square\clight}

\usepackage{cite} 
\usepackage{mciteplus}

\usepackage{longtable} 
\usepackage{todonotes}

\begin{document}

\renewcommand{\thefootnote}{\fnsymbol{footnote}}
\setcounter{footnote}{1}


\begin{titlepage}
\pagenumbering{roman}

\vspace*{-1.5cm}
\centerline{\large EUROPEAN ORGANIZATION FOR NUCLEAR RESEARCH (CERN)}
\vspace*{1.5cm}
\noindent
\begin{tabular*}{\linewidth}{lc@{\extracolsep{\fill}}r@{\extracolsep{0pt}}}
\ifthenelse{\boolean{pdflatex}}
{\vspace*{-2.7cm}\mbox{\!\!\!\includegraphics[width=.14\textwidth]{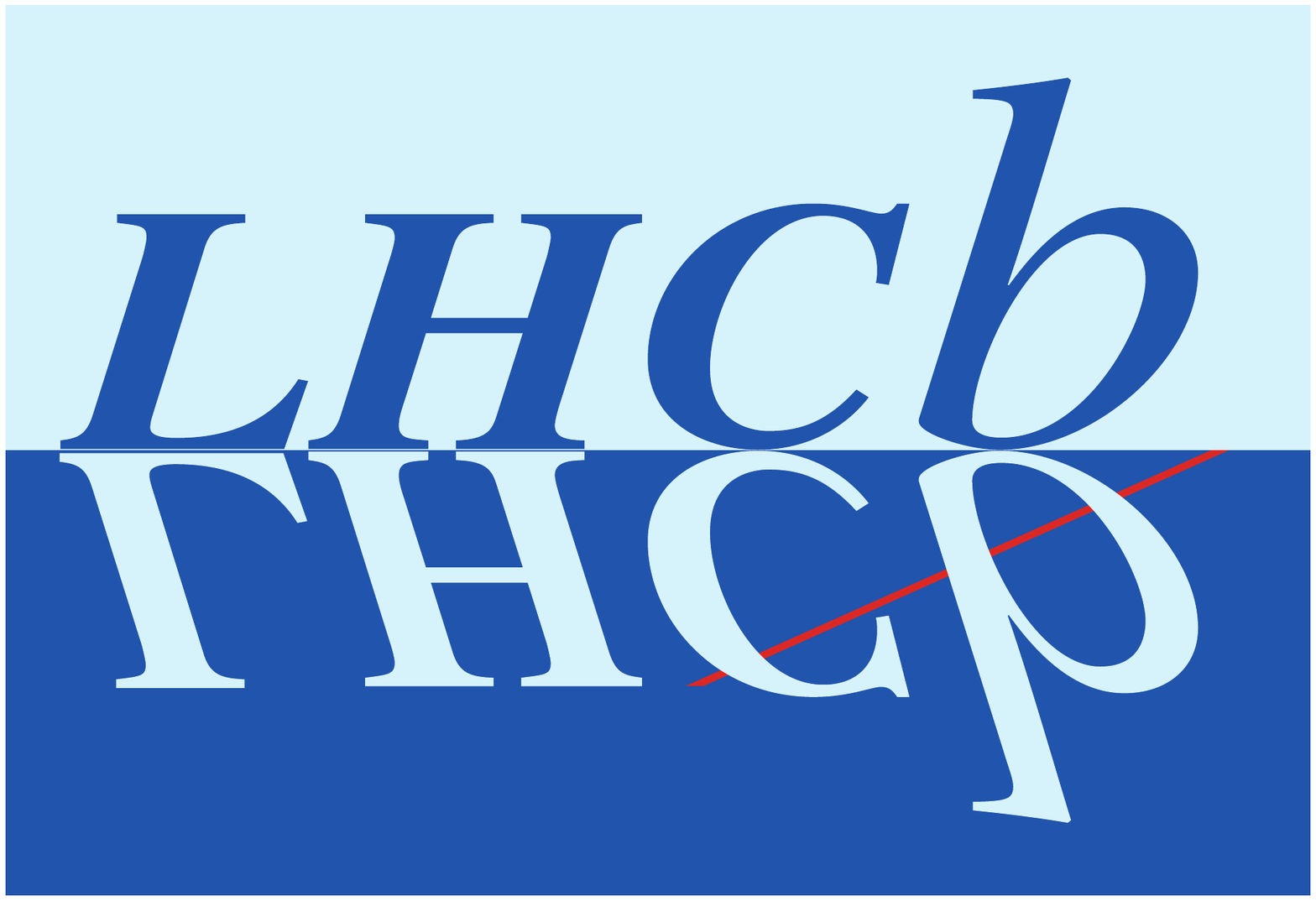}} & &}%
{\vspace*{-1.2cm}\mbox{\!\!\!\includegraphics[width=.12\textwidth]{lhcb-logo.eps}} & &}%
\\
 & & CERN-EP-2017-209 \\  
 & & LHCb-PAPER-2017-029\\  
 & & 27 November 2017 \\ 
 & & \\
\end{tabular*}

\vspace*{1.0cm} 

{\normalfont\bfseries\boldmath\huge
\begin{center}
  \papertitle
\end{center}
}

\vspace*{0.7cm}

\begin{center}
\paperauthors\footnote{Authors are listed at the end of this article.}
\end{center}

\vspace{\fill} 

\begin{abstract}
  \noindent
  A measurement is presented of decay-time-dependent \CP violation in the decays \mbox{\BdToJpsiKS} and \mbox{\BdToPsiTwoSKS}, where the $\Jpsi$ is reconstructed from two electrons and the $\psitwoS$ from two muons. The analysis uses a sample of $pp$ collision data recorded
  with the \lhcb experiment at centre-of-mass energies of \SI{7}{} and \SI{8}{\TeV}, corresponding to an integrated luminosity of
  $\SI{3}{\fb^{-1}}$. The \CP-violation observables are measured to be
  \begin{equation*}
  \begin{aligned}
  C\!\left(\BdToJpsiKS\right)          &= \phantom{+}\,0.12\,\pm\,0.07\pm\,0.02\,,\\
  S\!\left(\BdToJpsiKS\right)          &= \phantom{+}\,0.83\,\pm\,0.08\pm\,0.01\,, \\
  C\!\left(\BdToPsiTwoSKS\right)         &= -\,0.05\,\pm\,0.10\pm\,0.01\,,\\
  S\!\left(\BdToPsiTwoSKS\right)         &= \phantom{+}\,0.84\,\pm\,0.10\pm\,0.01\,,\\
  \end{aligned}
  \end{equation*}
  where $C$ describes \CP violation in the direct decay, and $S$ describes $\CP$
  violation in the interference between the amplitudes for the direct decay and
  for the decay after $\Bz$--$\Bzb$ oscillation. The first uncertainties are
  statistical and the second are systematic. The two sets of results are
  compatible with the previous \lhcb measurement using
  \BdToJpsiKS decays, where the $\Jpsi$ meson was reconstructed from two muons. The averages of all three sets of LHCb results are
  \begin{equation*}
  \begin{aligned}
  C(\Bz\to[\ccbar]\KS) &= -0.017 \pm 0.029\,,\\
  S(\Bz\to[\ccbar]\KS) &= \phantom{+}0.760 \pm 0.034\,,\\
  \end{aligned}
  \end{equation*}
  under the assumption that higher-order contributions to the decay amplitudes
  are negligible. The uncertainties include statistical and
  systematic contributions.  
\end{abstract}

\vspace*{1.0cm}

\begin{center}
  Published in JHEP 11 (2017) 170
\end{center}

\vspace{\fill}

{\footnotesize
\centerline{\copyright~\papercopyright, licence \href{\paperlicenceurl}{\paperlicence}.}}
\vspace*{2mm}

\end{titlepage}


\newpage
\setcounter{page}{2}
\mbox{~}

\cleardoublepage


\renewcommand{\thefootnote}{\arabic{footnote}}
\setcounter{footnote}{0}



\pagestyle{plain} 
\setcounter{page}{1}
\pagenumbering{arabic}


%


\section{Introduction}
\label{sec:introduction}
Precision measurements of $\CP$ violation in the decays of neutral $B$ mesons
provide stringent tests of the quark sector of the Standard Model (SM),
in which $\CP$ violation arises due to a single irreducible phase of the
Cabibbo-Kobayashi-Maskawa (CKM)
quark-mixing matrix~\cite{Cabibbo,KobayashiMaskawa}. The
\mbox{$\Bz\to[\ccbar]\KS$} family of decay modes, where $[\ccbar]$ denotes a
charmonium resonance ($\jpsi$, $\psitwos$, $\etac$, \etc), is ideal for studying
$\CP$ violation~\cite{Bigi:1981qs,Carter:1981}. Such decays proceed via a
$\bquark\to[\ccbar]\squark$ transition, where higher-order contributions that
could introduce additional strong and weak phases in the decay amplitudes are
expected to be small~\cite{Faller:1124258,Jung:P2012a,Frings:2015eva}. As $\Bz$
and $\Bzb$ mesons decay into a common final state in $\Bz\to[\ccbar]\KS$
decays,\footnote{The inclusion of charge-conjugate processes is implied
throughout the article, unless otherwise noted. The notation $\Bz$ refers to a
neutral $B$ meson containing a $\bquarkbar$ and a $\dquark$ quark including the
charge-conjugate state.} the interference between the direct decay and decay
after $\Bz$--$\Bzb$ mixing induces $\CP$ violation.

Since \CP violation in the mixing is known to be negligible~\cite{HFLAV16}, the
decay-time- and flavour-dependent decay rate for $\Bz$ and $\Bzb$ mesons can be
expressed as
\begin{equation}
\label{eq:decay_rate}
 \Gamma(t,d)\propto
  \mathrm{e}^{-\frac{t}{\tau}} \Big[
    \cosh(\DG\, t/2) + A_{\DG} \sinh(\DG\, t/2)
    - d\cdot S \sin(\dm\,t) + d\cdot C \cos(\dm\,t)
    \Big]\,,
\end{equation}
where in the equation the symbols are as follows: $t$ is the proper decay time;
$\tau$ is the mean lifetime of the $\Bz$ and $\Bzb$ meson; $\dm$ and $\DG$ are
the mass and decay width differences of the two \Bz mass eigenstates; $d$
represents the $\Bz$ meson flavour at production and takes values of
$+1$/$-1$ for mesons with an initial flavour of $\Bz$/$\Bzb$; and $S$, $C$, and
$A_{\DG}$ are the $\CP$-violation observables. The asymmetry between the $\Bzb$
and $\Bz$ decay rates is given by
\begin{equation}
\label{eq:signal_asymmetry}
\begin{split}
  {\cal A}_{[\ccbar]\KS}(t) & \equiv
    \frac{\Gamma(\Bzb(t)\!\rightarrow\![\ccbar]\KS)  - \Gamma(\Bz(t)\!\rightarrow\![\ccbar]\KS)}
         {\Gamma(\Bzb(t)\!\rightarrow\![\ccbar]\KS)  + \Gamma(\Bz(t)\!\rightarrow\![\ccbar]\KS)}\\
         &=\frac{S \sin(\dm\,t) - C \cos(\dm\,t)}
         {\cosh(\DG\,t/2) + A_{\DG} \sinh(\DG\,t/2)}\approx S \sin(\dm\,t) - C \cos(\dm\,t)\,,
\end{split}
\end{equation} where the approximate expression is valid under the assumption
$\DG=0$, which is well motivated at the current experimental
precision~\cite{HFLAV16}. The observable $C$ is related to $\CP$ violation in
the direct decay, while the observable $S$ corresponds to $\CP$ violation in the
interference. The world average of $C=-0.004\pm0.015$ as given by the Heavy
Flavor Averaging Group~\cite{HFLAV16} is compatible with zero. The observable
$S$ can be written as a function of one of the angles of the unitarity triangle
of the CKM matrix,
\mbox{$\beta\ \equiv\text{arg}\left[-\left(\Vcdnew\Vcbs\right)/\left(\Vtdnew\Vtbs\right)\right]$}, which is the most
precisely measured angle in the unitary triangle. In the limit of negligible higher-order
contributions, which is assumed when combining results from different
\mbox{$\Bz\to[\ccbar]\KS$} modes, $S$ can be identified as $\sintwobeta$.

Applying CKM unitarity and using measurements of other CKM-related quantities
leads to a SM prediction of $\sintwobeta =0.740\,^{+0.020}_{-0.025}$ by the
CKMfitter group~\cite{CKMfitter2015} and of $\sintwobeta = 0.724 \pm 0.028$ by the
UTfit collaboration~\cite{UTfit-UT}. The Belle and BaBar collaborations have
already constrained $\sintwobeta$ to a high precision in the
\mbox{$\BdToJpsiKS$} mode. They reported $S =0.670 \pm 0.032$~\cite{Adachi:2012et} and $S = 0.657 \pm 0.038$~\cite{Aubert:2009aw}, respectively.
The LHCb collaboration performed a measurement using $\Bz\to\jpsi\KS$
decays, where $\Jpsi$ meson was reconstructed from two muons, and obtained a value of $S = 0.73 \pm 0.04$~\cite{LHCb-PAPER-2015-004}.

This article presents a study of decay-time-dependent \CP violation in the
decays $\Bz\to\jpsi\KS$ and $\Bz\to\psitwoS\KS$ using data
collected with the \lhcb experiment in $pp$ collisions at centre-of-mass
energies of $7$ and $\SI{8}{\TeV}$, corresponding to a total integrated
luminosity of $\SI{3}{\fb^{-1}}$. In both decays, only the
$\pip\pim$ final state of the $\KS$ meson is considered. The $\Jpsi$ meson is
reconstructed from two electrons, whereas the $\psitwoS$ is reconstructed from
two muons. This is the first decay-time-dependent measurement at a hadron
collider that uses electrons in the final state. Including these additional
\mbox{$\Bz\to[\ccbar]\KS$} decay modes results in a $\SI{20}{\percent}$
improvement in the precision on \sintwobeta at \lhcb.


\section{Detector and event selection}
\label{sec:detector}

The \lhcb detector~\cite{Alves:2008zz,LHCb-DP-2014-002} is a single-arm forward
spectrometer covering the \mbox{pseudorapidity} range from 2 to 5, designed for
the study of particles containing \bquark or \cquark quarks. The detector
includes a high-precision tracking system consisting of a silicon-strip vertex
detector surrounding the $pp$ interaction region, a large-area silicon-strip
detector located upstream of a dipole magnet with a bending power of about
$4{\mathrm{\,Tm}}$, and three stations of silicon-strip detectors and straw
drift tubes placed downstream of the magnet. The tracking system provides a
measurement of momentum, \ptot, of charged particles with a relative uncertainty
that varies from 0.5\% at low momentum to 1.0\% at 200\gevc. The minimum
distance of a track to a primary vertex, PV, the impact parameter, IP, is
measured with a resolution of $(15+29/\pt)\mum$, where \pt is the component of
the momentum transverse to the beam, in\,\gevc. Different types of charged
hadrons are distinguished using information from two ring-imaging Cherenkov
detectors. Photons, electrons, and hadrons are identified by a calorimeter
system consisting of scintillating-pad and preshower detectors, an
electromagnetic calorimeter, and a hadronic calorimeter. As bremsstrahlung from
the electrons can significantly affect their momenta, a correction is applied
using the measured momenta of photons associated to the electron. Muons are
identified by a system composed of alternating layers of iron and multiwire
proportional chambers.

The online event selection is performed by a trigger, which consists of a
hardware stage, based on information from the calorimeter and muon systems,
followed by a software stage, which applies a full event reconstruction. In
the offline selection, trigger signals are associated with reconstructed
particles. Selection requirements can therefore be made on the trigger
selection itself and on whether the decision was due to the signal candidate,
other particles produced in the $pp$ collision, or a combination of both.
While in the case of the \jpsi mode an inclusive approach is chosen to keep
any candidate that passes both trigger stages, in the \psitwoS mode the muons
can be used in the decision of the trigger due to their clean signature in the
detector. For the $\psitwoS$ mode events are selected at the hardware stage
that contain at least one muon with transverse momentum $\pt>1.48\gevc$ in the
7\tev data or $\pt>1.76\gevc$ in the 8\tev data. In the subsequent software
stage events are required to contain either at least one muon with a
transverse momentum $\pt>1.0\gevc$ and $\mathrm{IP}>100\mum$ with respect to
all PVs in the event, or two oppositely charged muons with combined mass
$m(\mu^{+}\mu^{-})>2.7\gevc$. Finally, the tracks of two muons are required
to form a vertex that is significantly displaced from the PVs.

The selection strategies are similar for $\BdToJpsiKS$ and $\BdToPsiTwoSKS$
candidates. The $\Bz$ candidates are reconstructed by combining charmonium and
$\KS$ candidates that form a common vertex. The charmonium candidates are formed
from two oppositely charged tracks identified as electrons or muons. The pairs
of tracks need to be of good quality and must form a vertex that is
significantly displaced from any primary vertex. The muon candidates are
required to have momenta \hbox{$p>\SI{8}{\GeVc}$} and transverse momenta
\hbox{$\pt>\SI{1}{\GeVc}$}, and the dimuon invariant mass is in
the range \mbox{$3626<m(\mu^{+}\mu^{-})<\SI{3746}{\MeVcc}$}. The electron
candidates are required to have a $\pt>\SI{500}{\MeVc}$ and
\mbox{$2300<m(e^{+}e^{-})<\SI{4000}{\MeVcc}$}, where a wider range compared to
the dimuon mode is chosen to account for the worse resolution due to
bremsstrahlung. The decay vertex of the $\KS$ candidates must be significantly
displaced from any PV, while the dipion invariant mass needs to be consistent
with the known
\KS mass \cite{PDG2017}.

The invariant mass of each $\Bz$ candidate is determined by a kinematic
fit~\cite{Hulsbergen:2005pu}, where the masses of the lepton and pion pairs are
constrained to the known charmonium and $\KS$ masses, respectively. The mass of
the $\Bz$ candidates is required to be in the range
\mbox{${5150}<m(\Jpsi\KS)<\SI{5600}{\MeVcc}$} or
\mbox{${5200}<m(\psitwoS\KS)<\SI{5450}{\MeVcc}$}. The reconstructed decay time of the $\Bz$
candidates, $t'$, is obtained from a separate fit that constrains the $\Bz$
candidate to originate from a PV. The $\Bz$ candidates are kept if they have
kinematic fits of a good quality, measured decay times in the range
${0.2}<t'<\SI{15}{\ps}$ and decay-time-uncertainty estimates
$\sigma_t<\SI{0.4}{\ps}$.

To suppress combinatorial background, a multivariate selection is applied for
each mode in which a boosted decision tree (BDT)~\cite{Scikit} is trained
using the AdaBoost boosting algorithm~\cite{AdaBoost}. The BDTs are trained
using simulated signal samples and background samples consisting of $\Bz$
candidates with invariant masses above the considered regions, \ie
\mbox{${5600}<m(\Jpsi\KS)<\SI{6000}{\MeVcc}$} or
\mbox{${5450}<m(\psitwoS\KS)<\SI{5500}{\MeVcc}$}. The BDTs exploit features
related to kinematic and topological properties of the decay, along with track-
and vertex-reconstruction qualities. The common BDT features of the two decay
modes are $\pt(\KS)$, $\pt([\ccbar])$, the $\chisq$ values of the kinematic
fits, and the minimum and maximum of $\log(\chisqip)$ for each pion and for each
lepton, where $\chisqip$ is defined as the increase in $\chisq$ when including the
track in the PV fit. In addition to the common variables, the BDT for the
$\jpsi$ mode includes $\pt(\Bz)$, $\chisqip(\Jpsi)$, $\chisqip(\KS)$, and the
$\Bz$-decay vertex-fit quality, $\chisqvtx(\Bz)$. The BDT for the
$\psitwoS$ mode includes $\chisqvtx(\KS)$, and the $\KS$ decay-time
significance, $t/\sigma_t(\KS)$. The requirements on the BDT responses are
chosen to maximise the expected sensitivity on the $\CP$ observable $S$.

To suppress possible contamination from $\Lb\to[\ccbar]\PLambda(p\pim)$ decays,
the dipion invariant mass is calculated under the $p\pi$ invariant mass
hypothesis. Candidates compatible with the known $\PLambda$ mass\cite{PDG2017}
are rejected. In the case of the $\Jpsi$ mode an additional
proton-identification veto is applied. Aside from irreducible
$\Bs\to[\ccbar]\KS$ components that are modelled in the invariant mass fit, no
other contributing peaking backgrounds are found.

Multiple combinations of $\Bz$ candidates and PVs can occur in one event. After
applying all selection criteria less than \SI{1}{\percent} and
\SI{1.7}{\percent} multiple candidates are observed in the \jpsi and \psitwoS
mode, respectively. Of these remaining multiple $(\Bz, \text{PV})$ pairs per
event, one is chosen randomly.


\section{Invariant mass fit}
\label{sec:massfit}

\begin{figure}[tb]
\begin{center}
  \includegraphics[width=0.495\textwidth]{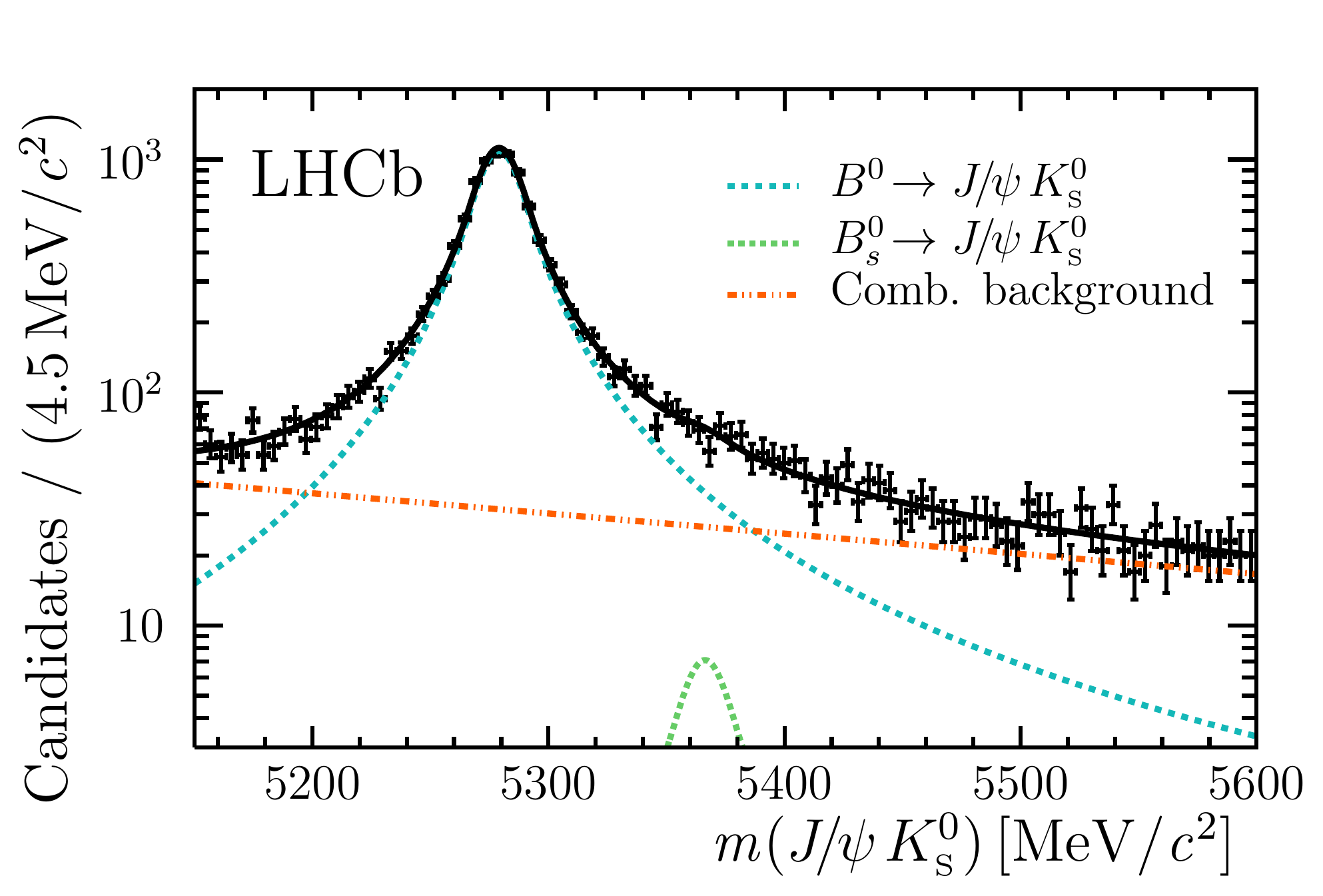}
  \includegraphics[width=0.495\textwidth]{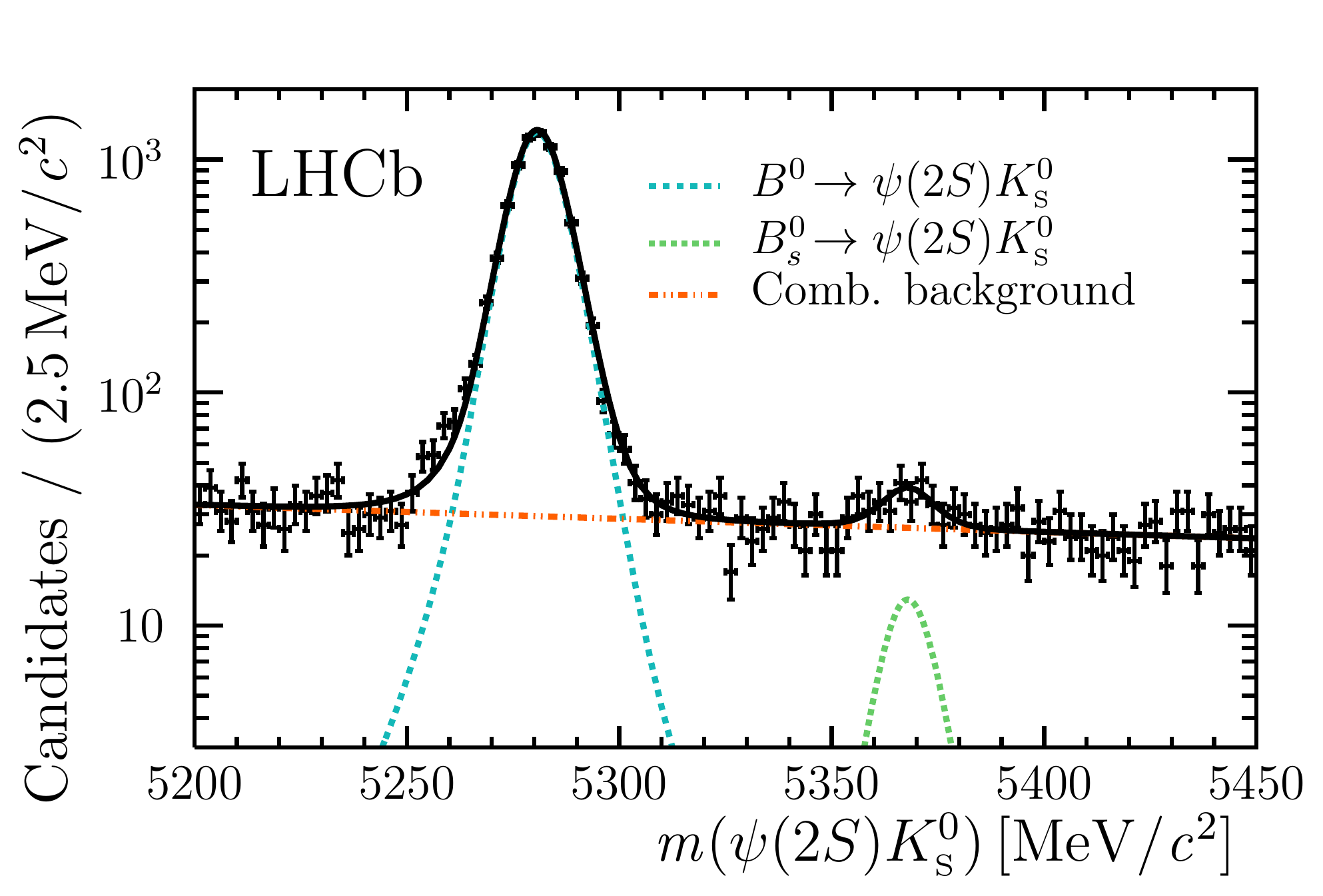}
  \caption{Invariant mass of the $\Bz$ candidates for (left) the \Jpsi and (right) the \psitwoS mode. The lines represent the result of the fit described in the text.}
    \label{fig:nominal_massfit}
\end{center}
\end{figure}

Unbinned maximum likelihood fits to the invariant mass distributions,
$m(\jpsi\KS)$ and $m(\psitwos\KS)$, are performed to determine signal candidate
weights using the \sPlot technique~\cite{Pivk:2004ty}. These signal candidate
weights are used to statistically subtract the background in the $\CP$ asymmetry
fit. The probability density functions (PDFs) of the signal and the
$\Bs\to[\ccbar]\KS$ background components are both parametrised by Hypatia
functions \cite{Santos:2014}, which consist of hyperbolic cores and power-law
tails. The values of the parameters describing the tails are taken from
simulation and used for both components. The widths of both components and the
mean of the \Bz component are allowed to vary in the fit, while the mean of the
\Bs component is offset from the \Bz mean by the known $\Bs$--$\Bz$ mass difference~\cite{PDG2017}.
The combinatorial background is described by an exponential function. The
invariant mass distributions and the fit results are shown in
Fig.\,\ref{fig:nominal_massfit}. The fits yield a total of $\SI{10630\pm140}{}$
\BdToJpsiKS decays and $\SI{7970\pm100}{}$ \BdToPsiTwoSKS decays with mass
resolutions of about $\SI{29}{\MeVcc}$ and $\SI{7}{\MeVcc}$, respectively. The
worse resolution for the \jpsi mode is caused by the energy loss of the final
state electrons, which cannot fully be corrected in the reconstruction.
%


\section{Flavour tagging}
\label{sec:tagging}

In a decay-time-dependent \CP-violation measurement, it is essential to know the
flavour of each $\Bz$ meson at production. Multiple flavour-tagging algorithms
are combined to achieve the best response. Each tagging algorithm
provides a decision (tag), $d' \in \{-1,0,1\}$, corresponding to a $\Bz$ candidate
tagged as $\Bzb$, untagged or tagged as $\Bz$, respectively, and the mistag
probability estimate, $\eta$. The tagging algorithms are categorised as
same-side, SS, and opposite-side,
OS~\cite{LHCb-PAPER-2011-027,LHCb-PAPER-2016-039,LHCb-PAPER-2015-027}. The SS
taggers exploit particles created in the fragmentation process of the
$\Bz$ meson, while the OS taggers use decay products of the accompanying
$\bquark$ hadron that is produced in association with the signal
$\Bz$ meson.

The combination of OS taggers used in this analysis is based on different possible final
states in the decay of the other $\bquark$ hadron in the event. The
tagging responses are determined from the charges of muons,
electrons or kaons; a weighted average of the charges of all tracks;
and the decay products of charm decays possibly originating from the other $\bquark$
hadron in the event. In the case of the SS taggers the tagging decision
is based on the charges of the pions and the protons originating from the
fragmentation process of the signal $\Bz$ mesons. The OS and SS decisions,
$\dOS$ and $\dSS$, and their mistag estimates, $\etaOS$ and $\etaSS$, are combined
for each $\Bz$ candidate. The tags $\dOS$ and $\dSS$ are combined event-by-event
during the fit procedure, taking into account their per-candidate mistag
estimates.

The mistag estimates are calibrated using flavour-specific channels that are
kinematically similar to the signal channels, so that $\eta$ on average matches
the signal mistag probability, $\omega(\eta)$. The difference in the tagging
response for $\Bz$ and $\Bzb$ mesons is taken into account. The calibration
channels are $\BuToJpsiK$ for the OS taggers, and $\BdToJpsiKst$ for the SS
taggers, where the $\Jpsi$ is either reconstructed from two electrons or two
muons. Selection criteria similar to the signal requirements are applied and
signal candidate weights to subtract backgrounds are determined by a fit to the
$\Bz$ invariant masses, $m([\ccbar]\Kp)$ and $m([\ccbar]\Kstarz)$, with the
\sPlot technique. Before calibrating the tagging
output the samples are weighted such that the relevant candidate kinematic
distributions and properties match those of the signal decay. These
distributions and properties are the pseudorapidity, the $\pt(\Bz)$, the number
of tracks and primary vertices, and the azimuthal angle.

The effective tagging efficiency,
\mbox{$\varepsilon_\text{eff}=\varepsilon_\text{tag}\langle\mathcal{D}(\eta)^2\rangle$}, is a measure of the effective statistical power of a data sample. Here,
$\varepsilon_\text{tag}$ is the tagging efficiency, defined as the
fraction of candidates with a nonzero tag decision, and $\langle
\mathcal{D}(\eta)^2\rangle$ is the effective dilution arising from the per-event
dilution $\mathcal{D}(\eta)\equiv1-2\omega(\eta)$. The effective tagging
efficiencies for the OS and SS taggers and their combination are listed in
Table\,\ref{tab:ft_performnace}. A higher effective tagging efficiency
in the \jpsi channel compared to the \psitwoS mode is observed for both the SS
and OS flavour tagging. While the SS taggers are positively affected by a higher
average $\pt(\Bz)$, the OS taggers benefit from the more inclusive trigger
strategy in the $\Jpsi$ mode leading to lower mistag probabilities as well as
higher tagging efficiencies in this mode.
\begin{table}[t] 
\centering
\caption{Effective flavour-tagging efficiencies in per cent of the SS and OS taggers and their combination.}
\begin{tabular}{ccc}
\toprule
Tagger    & \BdToJpsiKS & \BdToPsiTwoSKS\\ \midrule
OS        & \num{3.60 \pm 0.13}                                 & \num{2.46 \pm 0.05}\\
SS        & \num{2.40 \pm 0.28}                                 & \num{1.07 \pm 0.08}\\
OS + SS   & \num{5.93 \pm 0.29}                                 & \num{3.42 \pm 0.09}\\
\bottomrule
\end{tabular}
\label{tab:ft_performnace}
\end{table}


\section{\texorpdfstring{$\boldsymbol{\CP}$}{CP} asymmetry fit}
\label{sec:cpfit}

The \CP observables are determined by using an unbinned weighted maximum
likelihood fit to the decay-time distributions for all \mbox{$\BdToJpsiKS$} and
\mbox{$\BdToPsiTwoSKS$} candidates. The signal candidate
weights are determined from the mass fits described previously and used to
subtract the background so that only the signal components need to be modelled.
The PDF, $\mathcal{P}(t', \vec{d'}\,|\sigma_t, \vec{\eta})$, describes the
measured $\Bz$ candidate decay time and tags, \mbox{$\vec{d'} =\left(\dOS,\dSS\right)$}. It also depends on the per-candidate
decay-time-uncertainty estimate, $\sigma_t$, and the mistag probability estimates,
\mbox{$\vec{\eta} = \left(\etaOS,\etaSS\right)$}. 

The fit is performed simultaneously in both decay modes, sharing the parameters
describing the \Bz system, \ie the $\Bz$ meson lifetime, $\tau$, and the mass
difference, $\dm$, but allowing for different \CP observables. The decay-time
distribution of the signal components,
$\mathcal{P}_{\CP}(t,\vec{d'}\,|\vec{\eta})$, is derived from
Eq.\,\ref{eq:decay_rate} considering the production asymmetry, $A_P$, between
\Bzb and \Bz mesons~\cite{LHCb-PAPER-2016-062}. Using a PDF,
$\mathcal{P}_{\mathrm{tag}}(\vec{d'}\,|d,\vec{\eta})$, which describes the
distribution of tags based on the true production flavour and taking into
account the mistag probability estimates and efficiencies, leads to
\begin{equation}
\resizebox{0.94\hsize}{!}{$\mathcal{P}_{\CP}(t,\vec{d'}\,|\vec{\eta})\propto\sum_{d}\mathcal{P}_{\mathrm{tag}}(\vec{d'}\,|d,\vec{\eta})\left[1-d\cdot A_\mathrm{P}\right]e^{-t/\tau}\lbrace 1-d\cdot S\sin\left(\dm t\right) + d\cdot C\cos\left(\dm t\right)\rbrace\,.$}
\end{equation}
The decay-time resolution is taken into account by convolving $\mathcal{P}_{\CP}$
with a resolution function, $\mathcal{R}(t'-t|\sigma_t)$. Furthermore, the decay-time distribution is
multiplied by a decay-time-dependent reconstruction efficiency function,
$\varepsilon(t')$, in order to take into account the distortion coming from the
event reconstruction and selection. These corrections lead to the experimental decay-time distribution
\begin{equation}
\mathcal{P}\left(t', \vec{d'}\,|\sigma_t, \vec{\eta}\right)\propto\left(\mathcal{P}_{\CP}(t,\vec{d'}\,|\vec{\eta})\otimes\mathcal{R}(t'-t|\sigma_t)\right)\times\varepsilon(t')\,.
\end{equation}
The resolution function is modelled by three Gaussian functions which describe
the deviation of $t'$ from $t$. The widths of two of these Gaussian functions
are linear functions of $\sigma_t$ and therefore vary for each candidate. The
means are shared by all three Gaussians. The third Gaussian describes the
proper-time resolution of candidates that have been associated with the wrong
PV. The parameters of the resolution model are determined from simulated events
and fixed in the fit, leading to effective single Gaussian resolutions of
$\SI{67}{fs}$ for the \jpsi mode and $\SI{48}{fs}$ for the \psitwoS mode for
correctly associated \Bz candidates. A small decay-time bias of $\SI{3}{fs}$ is
observed in the simulation. This bias is neglected in the fit but is considered
as a source of systematic uncertainty. The decay-time-dependent efficiency
function is parametrized using cubic B-splines. The positions and the number of
the knots for the splines are optimized on simulated data, whereas the
coefficients are free fit parameters.

\begin{figure}[tb]
\begin{center}
  \includegraphics[width=0.495\textwidth]{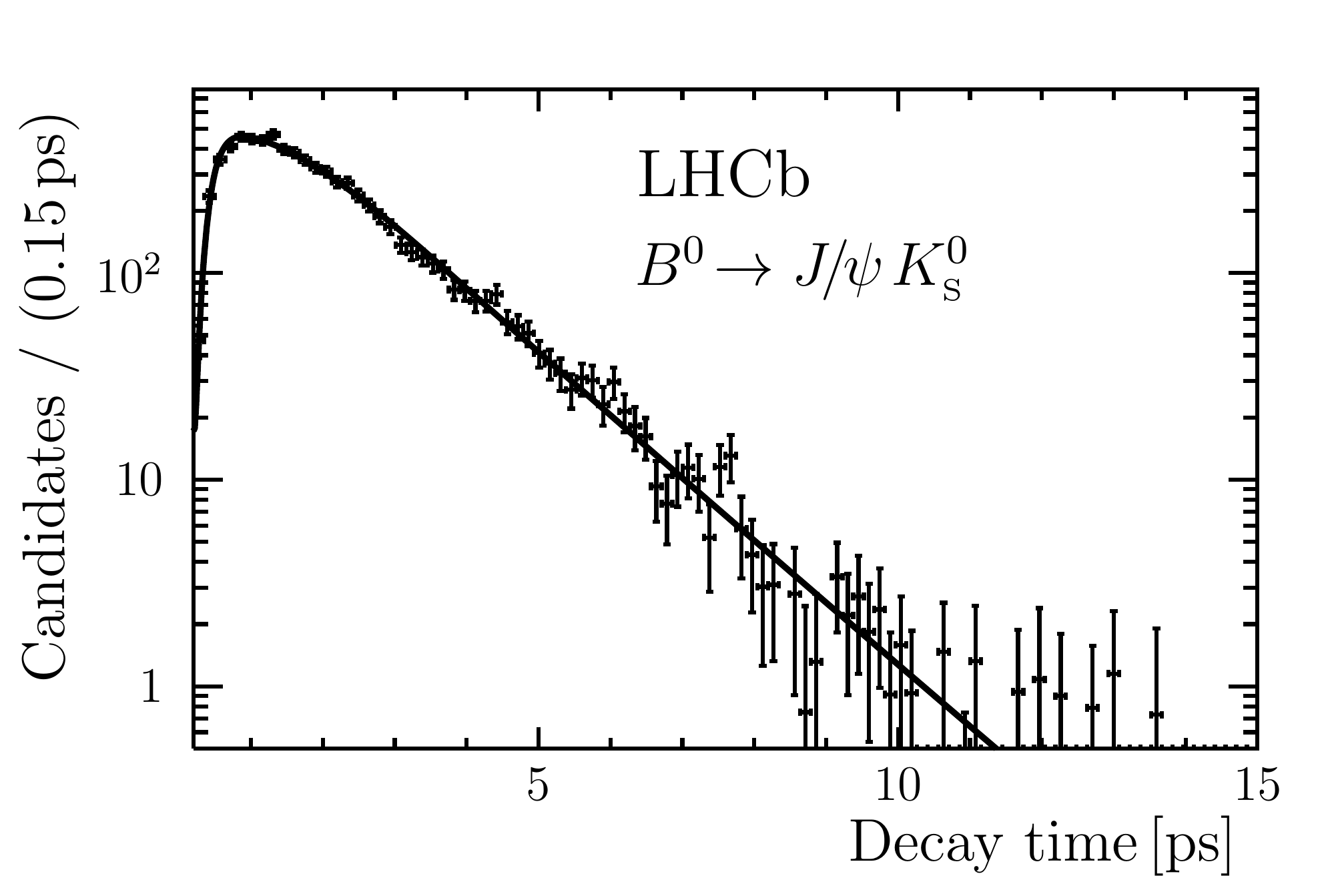}
  \includegraphics[width=0.495\textwidth]{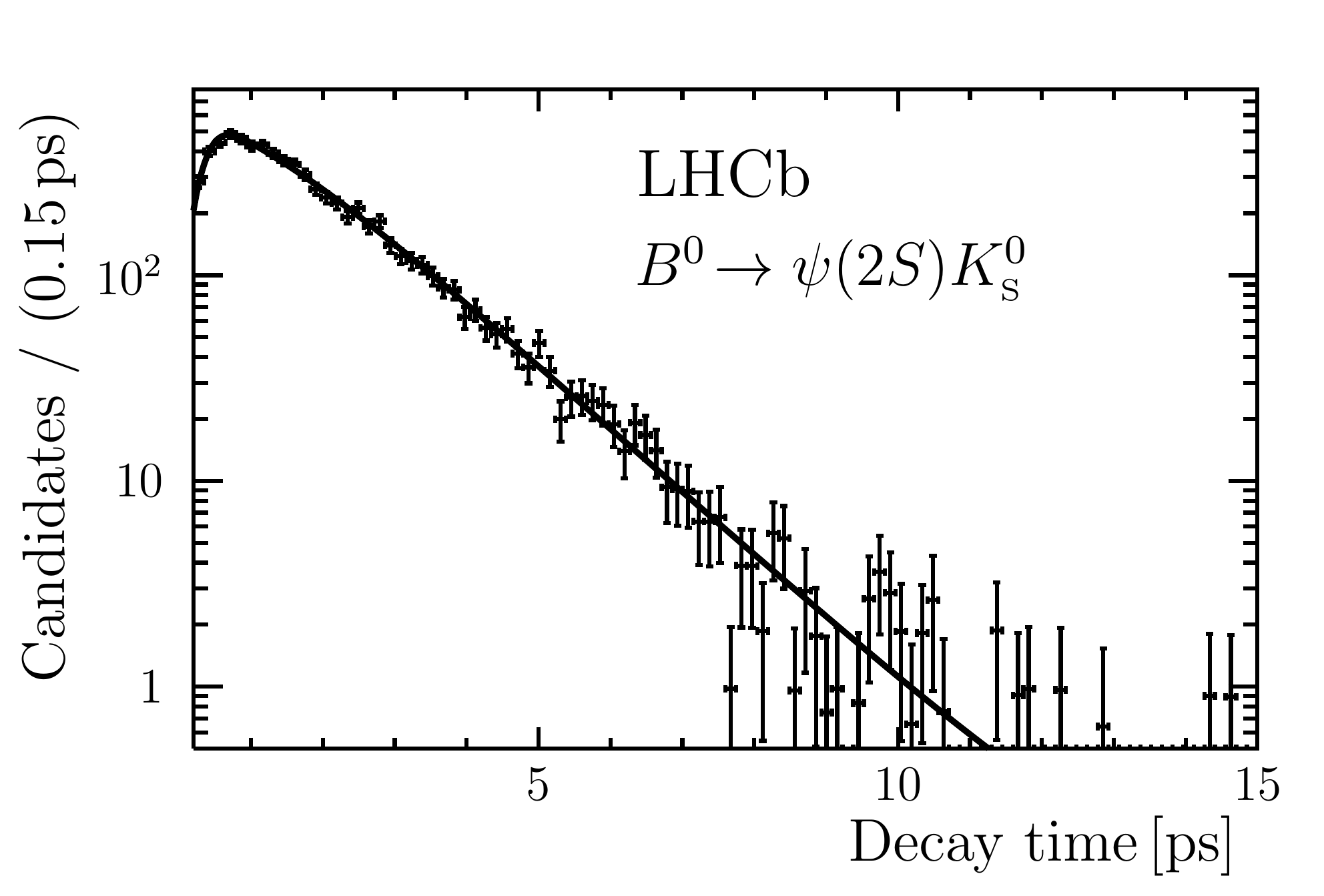}
  \caption{Projections of the decay-time fit to weighted (left) \mbox{\BdToJpsiKS} and \hspace{\textwidth} (right) \mbox{\BdToPsiTwoSKS} candidates.}
     \label{fig:results:fitprojection_jpsiee}
\end{center}
\end{figure}

Potential differences between simulation and data are accounted for as
systematic uncertainties. Production asymmetry values are evaluated for each
mode and centre-of-mass energy, using the recent LHCb measurement~\cite{LHCb-PAPER-2016-062} in bins of $\pt$ and rapidity of the
\Bz candidate. The values and uncertainties for the production asymmetry as well
as for the external inputs for the \Bz system are listed in
Table\,\ref{tab:decaytimefit:pars_constrained}. To propagate the uncertainties
in the fit, these parameters and also the tagging-calibration parameters are
Gaussian constrained using their statistical experimental uncertainties. Their
systematic uncertainties, as well as the uncertainty due to the assumption
$\DG=0$, are accounted for in the systematic studies. Tagging-calibration
parameters are constrained, taking into consideration their correlations. A fit
validation using pseudoexperiments is performed, showing no bias and correctly
estimated uncertainties from scans of the likelihood function~\cite{minuit}. The
reconstructed decay-time distributions and the corresponding fit projections are
shown in Fig.\,\ref{fig:results:fitprojection_jpsiee}.
%
\begin{table}[tb]
\caption{Parameters used as external inputs in the decay-time-dependent fit. The
production asymmetries are evaluated individually for both decay modes and separately for the different centre-of-mass energies of $7$ and $\SI{8}{\TeV}$. If two uncertainties are given the first is statistical and the second systematic. If one uncertainty is given it includes statistical and systematic contributions.}
\label{tab:decaytimefit:pars_constrained}
\centering
\begin{tabular}{lll}
  \toprule
  Parameter                     & Value and uncertainty & Source \\
  \midrule
  $\prodasym{\SI{7}{\TeV}}(\Jpsi)$               & $-0.0100 \pm 0.0084 \pm 0.0005$  &  \cite{LHCb-PAPER-2016-062} \\
  $\prodasym{\SI{8}{\TeV}}(\Jpsi)$               & $-0.0077 \pm 0.0054 \pm 0.0004$  &  \cite{LHCb-PAPER-2016-062} \\
  $\prodasym{\SI{7}{\TeV}}(\psitwoS)$            & $-0.0143 \pm 0.0077 \pm 0.0005$  &  \cite{LHCb-PAPER-2016-062} \\
  $\prodasym{\SI{8}{\TeV}}(\psitwoS)$            & $-0.0138 \pm 0.0051 \pm 0.0003$  &  \cite{LHCb-PAPER-2016-062} \\
  $\dm \left[\si{\invps}\right]$       & $\phantom{+}0.5065 \pm 0.0016 \pm 0.0011$  &  \cite{HFLAV16} \\
  $\tau [\si{\ps}]$                    & $\,\,\,\phantom{+}1.520  \pm 0.004\phantom{0}$                & \cite{HFLAV16} \\
  \bottomrule
\end{tabular}
\end{table}


\section{Systematic uncertainties}
\label{sec:sytematics}
Systematic uncertainties arise due to possible mismodelling of the PDFs and from
the uncertainties on the external inputs. The corresponding effects
are studied using simulated pseudoexperiments in which ensembles are generated
using parameters that differ from those used in the nominal fit. The
generated datasets are then fitted with the nominal model to test whether biases
in the parameters of interest occur.

The effect of neglecting \DG in the nominal model is studied by varying its
value within one standard deviation of its current experimental
uncertainty~\cite{HFLAV16}. Effects coming from the constrained inputs are
evaluated by varying their values by one standard deviation in terms of their
systematic experimental uncertainties. The constrained inputs are the production
asymmetry parameters, the oscillation frequency, $\dm$, the lifetime, $\tau$, as
well as the tagging-calibration parameters. The systematic uncertainty
arising due to the decay-time bias is evaluated using pseudoexperiments in which
a corresponding value of $\SI{3}{fs}$ is assumed. Furthermore, deviations in the
scaling of $\sigma_t$ are estimated at the level of $\pm\SI{30}{\%}$ and
addressed through varying the corresponding factors by this amount. Possible
inaccuracies in the decay-time-reconstruction efficiency are studied using a
different parameterization obtained from data. Table \ref{tab:syst_total}
summarizes the results of these studies. The individual uncertainties are added
in quadrature to obtain the overall systematic uncertainties.

The fit results are corrected for $\CP$ violation in $\Kz$--$\Kzb$ mixing and for
the difference in the nuclear cross-sections in material between \Kz and \Kzb interactions
\cite{PhysRevD.84.111501}. The numerical values of these corrections are
${-0.003}$ (${-0.004}$) for $S$ and ${+0.002}$ (${+0.002}$) for $C$ in the
\jpsi (\psitwoS) mode.
\begin{table}[tb]
\caption{Systematic uncertainties for the \CP-violation observables $S$ and $C$.}
\label{tab:syst_total}
  \centering
    \begin{tabular}{lcccc}
      \toprule
                & \multicolumn{2}{c}{\BdToJpsiKS} & \multicolumn{2}{c}{\BdToPsiTwoSKS} \\
      Source    & {$\sigma_S$} & {$\sigma_C$} & {$\sigma_S$} & {$\sigma_C$}    \\
      \midrule
      \small{$\DG$}                   & 0.003  & 0.007  & 0.007  & 0.003 \\
      \small{$\dm$}                   & 0.004  & 0.004  & 0.004  & 0.004 \\
      \small{Production asymmetry}    & 0.004  & 0.009  & 0.007  & 0.005 \\
      \small{Tagging calibration}     & 0.002  & 0.005  & 0.005  & 0.002 \\
      \small{Decay-time bias}         & 0.006  & 0.006  & 0.006  & 0.004 \\
      \small{$\sigma_t$ scaling}      & 0.003  & 0.005  & 0.002  & 0.002 \\
      \small{Decay-time efficiency}   & 0.006  & 0.004  & 0.006  & 0.004 \\
      \midrule
      Total                           & 0.011 & 0.016 & 0.014 & 0.010 \\
      \bottomrule
    \end{tabular}
\end{table}


\section{Results and conclusion}
\label{sec:results}

\begin{figure}[tb]
\begin{center}
  \includegraphics[width=0.495\textwidth]{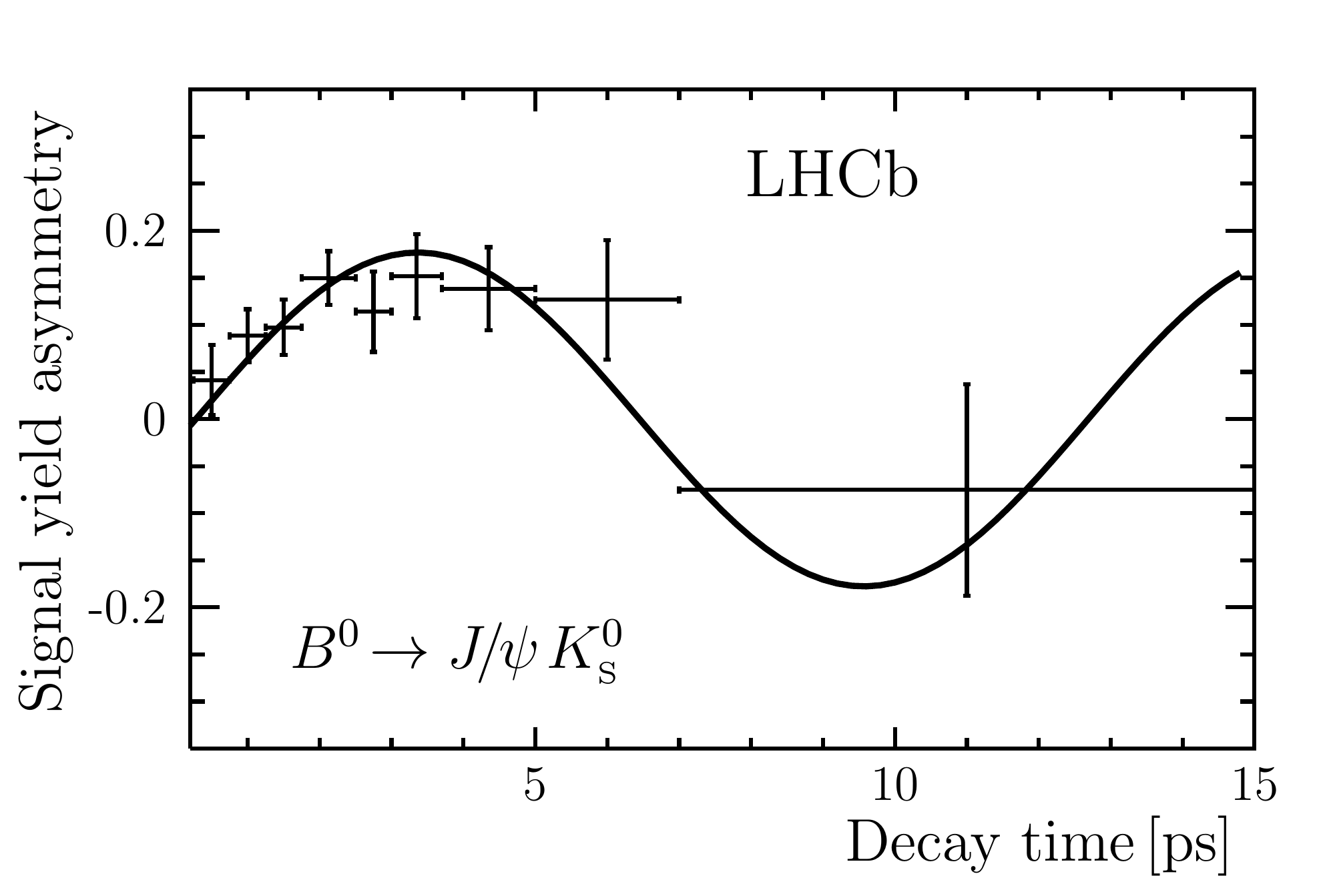}
  \includegraphics[width=0.495\textwidth]{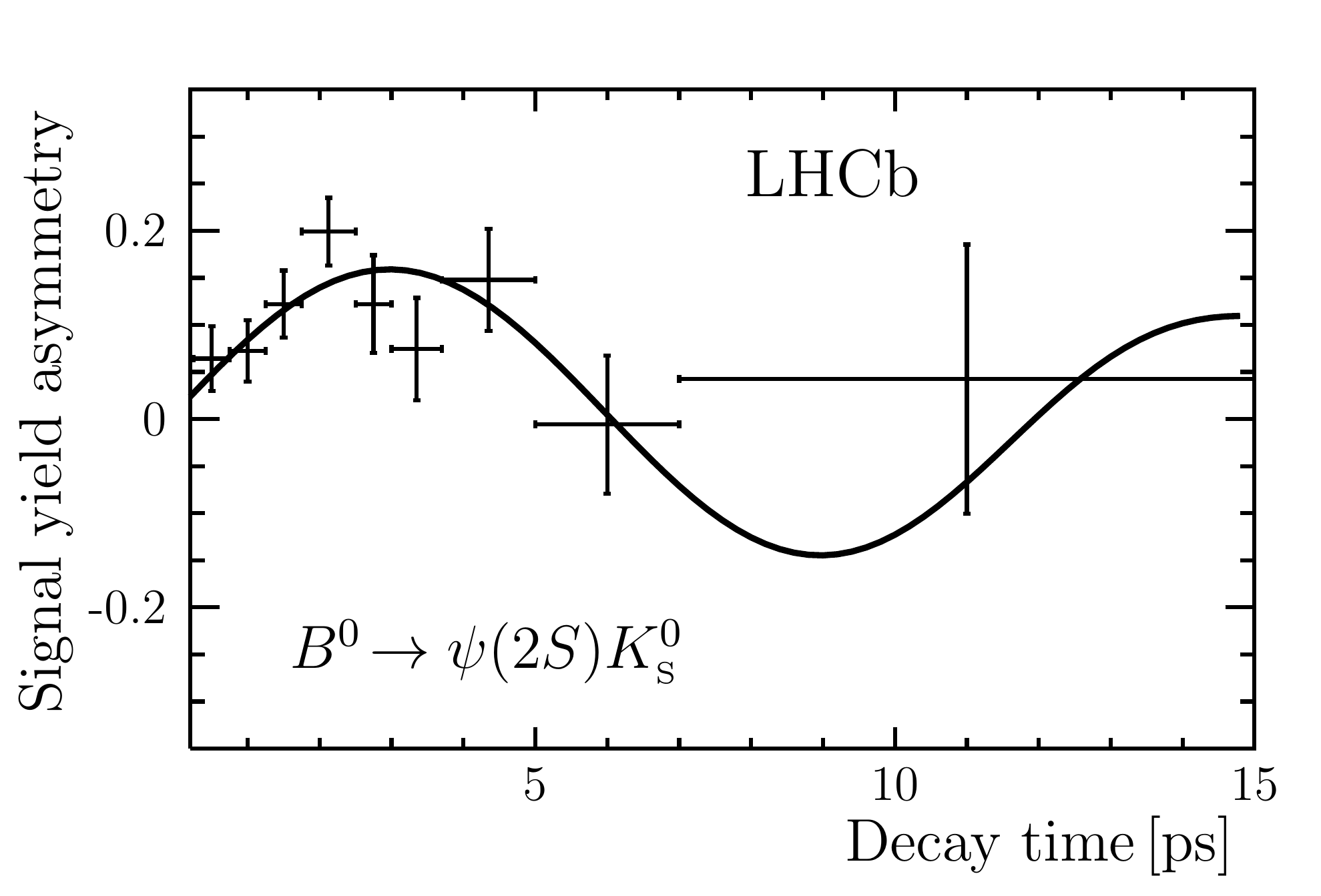}
  \caption{Signal yield asymmetries $(N_\Bzb - N_\Bz)/(N_\Bzb + N_\Bz)$ versus the decay time for \hspace{\textwidth} (left) \BdToJpsiKS and (right) \BdToPsiTwoSKS decays. The symbol $N_\Bz$ ($N_\Bzb$) is the number of decays with a $\Bz$ ($\Bzb$) flavour tag. The solid curves are the projections of the PDF with the combined flavour tagging decision.}
  \label{fig:asymmetry}
\end{center}
\end{figure}

The analysis of $\SI{10630\pm140}{}$ \BdToJpsiKS and $\SI{7970\pm100}{}$
\BdToPsiTwoSKS decays, where the $\Jpsi$ is reconstructed from two electrons and
the $\psitwoS$ from two muons, in a sample corresponding to $\SI{3}{\fb^{-1}}$ of
$pp$ collision data results in the \CP-violation observables
\begin{equation*}
\begin{aligned}
C\!\left(\BdToJpsiKS\right)          &= \phantom{+}\,0.12\,\pm\,0.07\pm\,0.02\,, \\
S\!\left(\BdToJpsiKS\right)          &= \phantom{+}\,0.83\,\pm\,0.08\pm\,0.01\,, \\
C\!\left(\BdToPsiTwoSKS\right)         &= -\,0.05\,\pm\,0.10\pm\,0.01\,,\\
S\!\left(\BdToPsiTwoSKS\right)         &= \phantom{+}\,0.84\,\pm\,0.10\pm\,0.01\,, \\
\end{aligned}
\end{equation*}
with correlation coefficients between $S$ and $C$ of $\num{0.46}$ and
$\num{0.48}$ for the \jpsi and the \psitwoS mode, respectively. The first
uncertainties are statistical and the second are systematic.
The signal yield asymmetries, $(N_\Bzb - N_\Bz)/(N_\Bzb + N_\Bz)$, as a function of decay time are shown in
Fig.\,\ref{fig:asymmetry}, where $N_\Bz$ ($N_\Bzb$) is the number of decays with a $\Bz$ ($\Bzb$) flavour tag.
The results for the electron and muon modes are compatible with each other and
with the previous \lhcb measurements using \BdToJpsiKS decays of \mbox{$S = 0.73
\pm 0.04$} and \mbox{$C= -0.038 \pm0.032$}~\cite{LHCb-PAPER-2015-004}, where the
$\Jpsi$ is reconstructed from two muons.

\begin{figure}[tb]
\begin{center}
  \includegraphics[width=0.495\textwidth]{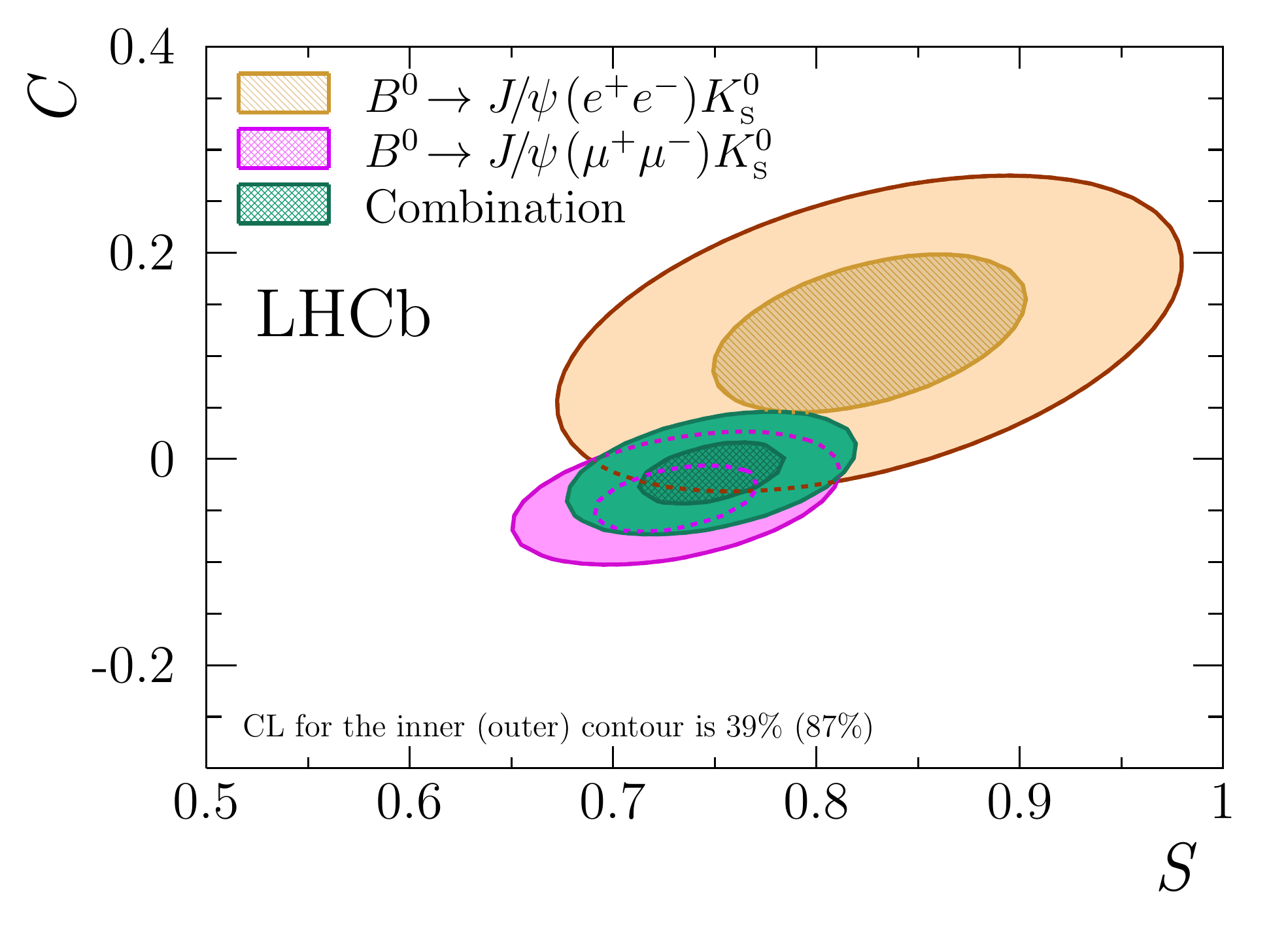}
  \includegraphics[width=0.495\textwidth]{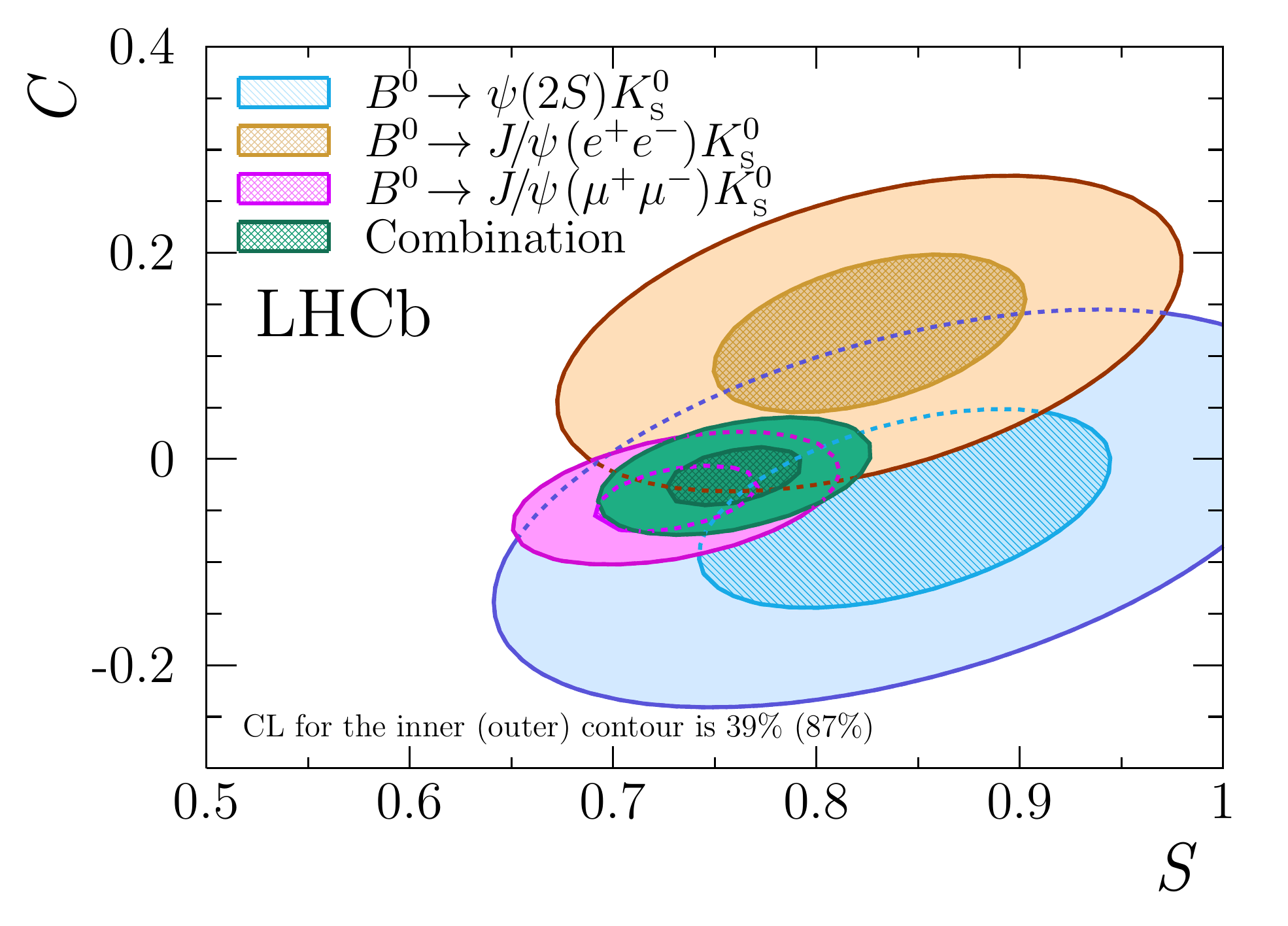}
  \caption{Two-dimensional likelihood scans for the combination of the (left) $\BdToJpsiKS$ modes and (right) all $\Bz\to[\ccbar]\KS$ modes. The confidence level for the inner (outer) contour is $39\%$ ($87\%$).}
    \label{fig:results:likelihoodscans}
\end{center}
\end{figure}

Combinations are performed using two-dimensional likelihood scans (see
Fig.\,\ref{fig:results:likelihoodscans}) taking into account the correlations
between the single measurements. The quoted uncertainties include statistical
and systematic contributions. Combining the \lhcb results for both $\jpsi$ modes
leads to
\begin{equation*}
\begin{aligned}
C(\BdToJpsiKS) &= -0.014 \pm 0.030\,,\\
S(\BdToJpsiKS) &= \phantom{+}0.75\phantom{0} \pm 0.04\,,\\
\end{aligned}
\end{equation*}
with a correlation coefficient of $\num{0.42}$. This combination is compatible
within $\num{1.9}$ standard deviations with the \BdToJpsiKS average of the
$B$-factories~\cite{HFLAV16}, while the result for the \psitwoS mode is
compatible within $\num{0.3}$ standard deviations with the \BdToPsiTwoSKSBfac
average of the $B$-factories~\cite{HFLAV16}. Building an \lhcb average using the
results from all $\Bz\to[\ccbar]\KS$ modes, \ie \BdToJpsiKS, where the $\Jpsi$
is either reconstructed from two muons or two electrons, and
\BdToPsiTwoSKS, the \CP-violation observables are determined to be
\begin{equation*}
\begin{aligned}
C(\Bz\to[\ccbar]\KS) &= -0.017 \pm 0.029\,,\\
S(\Bz\to[\ccbar]\KS) &= \phantom{+}0.760 \pm 0.034\,,\\
\end{aligned}
\end{equation*}
with a correlation coefficient of $\num{0.42}$. These results are consistent
with indirect measurements by the CKMfitter group~\cite{CKMfitter2015} and the UTfit
collaboration~\cite{UTfit-UT}. Furthermore, they improve the precision of $\sintwobeta$ at
\lhcb by $\SI{20}{\percent}$, and are expected to improve the precision of the
world average.
\clearpage

\section*{Acknowledgements}
%
%
\noindent We express our gratitude to our colleagues in the CERN
accelerator departments for the excellent performance of the LHC. We
thank the technical and administrative staff at the LHCb
institutes. We acknowledge support from CERN and from the national
agencies: CAPES, CNPq, FAPERJ and FINEP (Brazil); MOST and NSFC
(China); CNRS/IN2P3 (France); BMBF, DFG and MPG (Germany); INFN
(Italy); NWO (The Netherlands); MNiSW and NCN (Poland); MEN/IFA
(Romania); MinES and FASO (Russia); MinECo (Spain); SNSF and SER
(Switzerland); NASU (Ukraine); STFC (United Kingdom); NSF (USA).  We
acknowledge the computing resources that are provided by CERN, IN2P3
(France), KIT and DESY (Germany), INFN (Italy), SURF (The
Netherlands), PIC (Spain), GridPP (United Kingdom), RRCKI and Yandex
LLC (Russia), CSCS (Switzerland), IFIN-HH (Romania), CBPF (Brazil),
PL-GRID (Poland) and OSC (USA). We are indebted to the communities
behind the multiple open-source software packages on which we depend.
Individual groups or members have received support from AvH Foundation
(Germany), EPLANET, Marie Sk\l{}odowska-Curie Actions and ERC
(European Union), ANR, Labex P2IO, ENIGMASS and OCEVU, and R\'{e}gion
Auvergne-Rh\^{o}ne-Alpes (France), RFBR and Yandex LLC (Russia), GVA,
XuntaGal and GENCAT (Spain), Herchel Smith Fund, the Royal Society,
the English-Speaking Union and the Leverhulme Trust (United Kingdom).



\setboolean{inbibliography}{true}
\bibliographystyle{LHCb}
\bibliography{LHCb-PAPER,LHCb-CONF,LHCb-DP,LHCb-TDR,bibliography,main}

\newpage
\centerline{\large\bf LHCb collaboration}
\begin{flushleft}
\small
R.~Aaij$^{40}$,
B.~Adeva$^{39}$,
M.~Adinolfi$^{48}$,
Z.~Ajaltouni$^{5}$,
S.~Akar$^{59}$,
J.~Albrecht$^{10}$,
F.~Alessio$^{40}$,
M.~Alexander$^{53}$,
A.~Alfonso~Albero$^{38}$,
S.~Ali$^{43}$
% ...

{\footnotesize \it
% ...
$ ^{5}$Clermont Universit{\'e}, Universit{\'e} Blaise Pascal, CNRS/IN2P3, LPC, Clermont-Ferrand, France\\
% ...
$ ^{10}$Fakult{\"a}t Physik, Technische Universit{\"a}t Dortmund, Dortmund, Germany\\
% ...
$ ^{38}$ICCUB, Universitat de Barcelona, Barcelona, Spain\\
$ ^{39}$Universidad de Santiago de Compostela, Santiago de Compostela, Spain\\
$ ^{40}$European Organization for Nuclear Research (CERN), Geneva, Switzerland\\
% ...
$ ^{43}$Nikhef National Institute for Subatomic Physics, Amsterdam, The Netherlands\\
% ...
$ ^{48}$H.H. Wills Physics Laboratory, University of Bristol, Bristol, United Kingdom\\
% ...
$ ^{53}$School of Physics and Astronomy, University of Glasgow, Glasgow, United Kingdom\\
% ...
$ ^{59}$University of Cincinnati, Cincinnati, OH, United States\\
% ...
}
\end{flushleft}

\end{document}